\documentclass[letter]{aa} % Letter
\usepackage{graphicx}
\usepackage{url}
\usepackage{txfonts}
\usepackage{natbib}
\bibpunct{(}{)}{;}{a}{}{,}
\begin{document}
   \title{Oscillations in ${\rm \beta}$\,UMi}

   \subtitle{Observations with SMEI}

   \author{N.~J.~Tarrant 
				\and
          W.~J.~Chaplin
				\and
          Y.~Elsworth
				\and
          S.~A.~Spreckley
			 				\and
          I.~R.~Stevens
          }

   \offprints{N.~J.~Tarrant\\
		\email{njt@bison.ph.bham.ac.uk}
	}

   \institute{	School of Physics and
							Astronomy, University of Birmingham, Edgbaston, Birmingham
							B15 2TT, U.K.\\
             }

   \date{}

% \abstract{}{}{}{}{} 
% 5 {} token are mandatory
 
  \abstract
   {}
   {
		From observations of the K4III star $\beta$ UMi we attempt to determine whether oscillations or any other form of variability is present.
	}
   {
		A high-quality photometric time series of $\approx$1000 days in length obtained from the SMEI instrument on the Coriolis satellite is analysed. Various statistical tests were performed to determine the significance of features seen in the power density spectrum of the light curve. 	}
   { 
		Two oscillations with frequencies 2.44 and 2.92 $\mu$Hz have been identified. We interpret these oscillations as consecutive overtones of an acoustic spectrum, implying a large frequency spacing of 0.48 $\mu$Hz. Using derived asteroseismic parameters in combination with known astrophysical parameters, we estimate the mass of $\beta$ UMi to be $1.3 \pm 0.3\ {\rm M}_{\odot}$. Peaks of the oscillations in the power density spectrum show width, implying that modes are stochastically excited and damped by convection. The mode lifetime is estimated at \hbox{$18\pm9$~days}.
	}
   {}


   \maketitle

%______________________________________________________________________

\section{Introduction}

In this Letter we present new results on the variability of the K4III giant ${\rm \beta}$ UMi (Kochab) from observations made by the Solar Mass Ejection Imager (SMEI) on board the Coriolis satellite. While the star is a suspected variable, as noted by its entry in the New Catalogue of Suspected Variable Stars \citep[NSV 6846; ][]{1981CSV...C......0K}, the period and amplitude of variability have not yet been presented in the literature. Because it is on the cool side of the instability strip, we may expect $\beta$ UMi to show Sun-like oscillations, that is, oscillations stochastically excited and damped by convective noise.

Stellar parameters on $\beta$ UMi, taken from the revised Hipparcos catalogue \citep{2007hnrr.book.....V}, \citet{2003AJ....126.2502M}, and \citet{2003A&A...400..709D}, are shown in Table \ref{Table:BetaUmiParams}. We used the scaling laws of \citet[][hereafter KB95]{1995A&A...293...87K} to make predictions for Sun-like oscillations of the star. The maximum power of the modes is expected to be in the range of 3.3 to 4.7 $\mu$Hz (Eq. 10, KB95), with a significant RMS amplitude of 1.1 to 2.1 parts per thousand (ppt) (Eq. 8, KB95). The `large frequency spacing' (between adjacent harmonics) is expected to be small at 0.6 to 0.8 $\mu$Hz (Eq. 9, KB95). This may present a problem if, as has been suggested by \citet{2006A&A...448..709S}, mode lifetimes in giants are short, meaning neighbouring modes may overlap in frequency; however, other recent determinations \citep{2007MNRAS.382L..48T} suggest mode lifetimes may be longer in some giants, which would alleviate such problems. 

\begin{table}

\centering

\caption{Stellar Parameters for ${\rm \beta}$\,UMi. }

\label{Table:BetaUmiParams}

\begin{tabular}{lr}

\hline\hline
Parameter		&	Value		\\
\hline
\textbf{Literature Values:} \\
Parallax \footnotemark[1] 	&	$ 24.91 \pm 0.12\ {\rm mas} $ \\
V$_{\rm mag}$ \footnotemark[1] 	&	$ 2.2044 \pm 0.0008 $	\\
B-V	\footnotemark[1]			&	$ 1.465	\pm 0.005 $	\\
Mass (from evolutionary tracks) \footnotemark[2]	&	$	2.2 \pm 0.3	\ {\rm M}_{\odot } $ \\
Mass (from $\log g$)	\footnotemark[2]	&	$	2.5 \pm 0.9	\ {\rm M}_{\odot } $ \\
Angular Diameter \footnotemark[3]	&	$ 10.3 \pm 0.1 \ {\rm mas}	$ \\
\hline
\textbf{Derived Values:} \\
Distance		&	$ 40.1 \pm 0.2 \ {\rm pc } $ \\
Effective Temperature	&	$	4040 \pm 100	\ {\rm K} $ \\
Luminosity		&	$ 475 \pm 30 \ {\rm L}_{\odot } $ \\
Radius (photometric) 
				&	$ 43.5 \pm 0.5 \ {\rm R}_{\odot } $ \\
Radius (interferometric)
				&	$ 44.4 \pm 0.7 \ {\rm R}_{\odot} $ \\
\hline
\textbf{Predictions:} \\
Frequency of maximum power, $\nu_{\rm max}$	&	$ 4.0 \pm 0.7 \ \mu{\rm Hz}	$ \\
Large frequency spacing, $\Delta \nu$	&	$ 0.7 \pm 0.1 \ \mu{\rm Hz}	$ \\
Maximum predicted RMS amplitude in SMEI, \\
$\delta L/L$ (700 nm)	&	$ 1.6 \pm 0.5 \ {\rm ppt}	$ \\
\hline\hline

\end{tabular}

\end{table}

\footnotetext[1]{from the revised Hipparcos catalogue \citep{2007hnrr.book.....V}}
\footnotetext[2]{from two determinations in \citet{2003A&A...400..709D}}
\footnotetext[3]{from \citet{2003AJ....126.2502M}}

The SMEI observations we present here extend over a period of around three years and have a good overall duty cycle. The excellent resolution in frequency has allowed us to identify two individual modes of oscillation in ${\rm \beta}$\,UMi, and to also place direct constraints on the mode lifetime. 

%______________________________________________________________________

\section{Data and light-curve extraction}

The reader is referred to \citet{2007MNRAS.382L..48T} for an overview of the processing of SMEI data for use in asteroseismology. A more detailed description of the data extraction and processing pipeline is given in Spreckley and Stevens (in prep).

In summary, the SMEI instrument consists of three cameras facing respectively along, perpendicular to, and away from the Earth-Sun direction. Each of these cameras images a 60-by-3 degree slice of the zenith-facing hemisphere at a 4-second cadence, resulting in data on a 170-by-3 degree slice of the sky. Due to the Coriolis satellite's orbital geometry, this slice will co-rotate with the satellite over the course of an orbit such that the entire sky is imaged at an approximately 100-minute cadence \citep{2004SoPh..225..177J}.

The individual 60-by-3 degree images are then processed. Images with high background are removed, bias and dark current corrections are performed, and images are then flat-fielded and cosmic rays removed. A stray light correction is also performed. Once the images have been cleaned, aperture photometry based upon a modified form of the DAOPHOT routines \citep{1987PASP...99..191S} is performed. As a final step, corrections are made for long-term degradation of the CCD, and a position-dependent correction is applied to compensate for variation in the point-spread function. A single photometric measurement of intensity is thus obtained once per orbit. 

As ${\rm \beta}$\,UMi is at high declination (+74$^{\circ}$), it is only observed by camera \#\,2, which points perpendicular to the Earth-Sun direction; however, the coverage was very good, with fill of approximately 70\%. As the data set contained a number of asymmetrically distributed outliers, a simple arithmetic mean would not provide a good estimate of the average flux.
For this reason a mean of the entire time series was taken, which was calculated from only those points lying within 3 times the median absolute deviation from the global median value ($\approx$98\% of all points).
The average flux thus determined was subtracted from each value, and the resulting deviations were divided through by the average to give each point as a fractional change in intensity, $\delta L / L$.
Point-to-point scatter in the time domain is $\approx$1.5~parts per thousand (ppt).

   \begin{figure}
   \centering
   \includegraphics[width=0.5\textwidth]{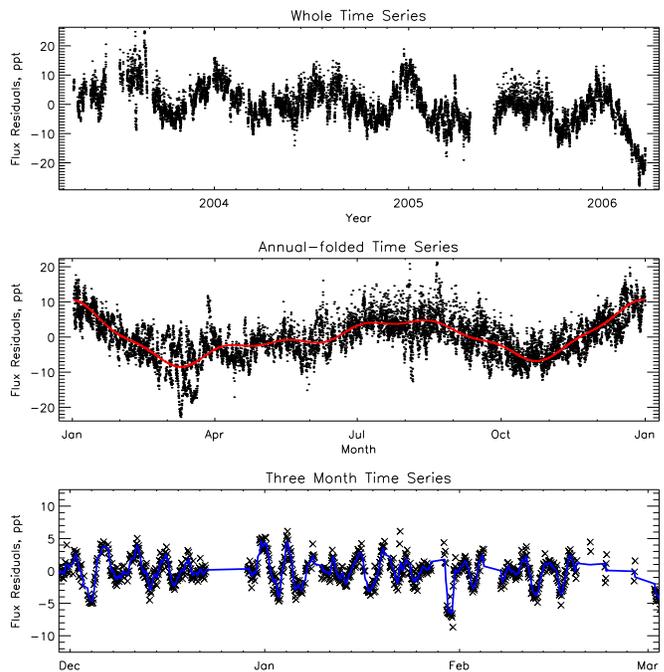}
      \caption{Data in the time domain - the whole time series (top), phase-folded across 365.26 days with the fitting to the annual cycles in red (middle), a three month time series with smoothed data (blue) (bottom). }
         \label{Fig:TimeDomain}
   \end{figure}

Figure \ref{Fig:TimeDomain} (top) shows the residual light curve of the time series in units of parts per thousand. A long-term trend can be readily observed in the data with a periodicity of around a year. These trends have been linked to an instrumental effect correlated to the instrument temperature. Work is on going to remove these trends in a later version of the data-generation pipeline. Due to this instrumental effect, it is difficult to draw conclusions regarding periodicities associated with stellar rotation, which would be expected to have a comparable period of a few hundred days. A phase-folded plot (Fig. \ref{Fig:TimeDomain}, middle) with a period of 1 year shows the annual and semi-annual cycles in the data, with amplitudes of a few ppt. Further harmonics of these cycles were also present at significant amplitude. Fits were performed up to the eighth harmonic of a year ($n=8$) and were removed from the data by subtraction in the time domain, leading to a reduction in the RMS amplitude of the time series from $\approx$6 to 4 ppt.

\begin{figure*}{}
   \sidecaption
   \includegraphics[width=12cm]{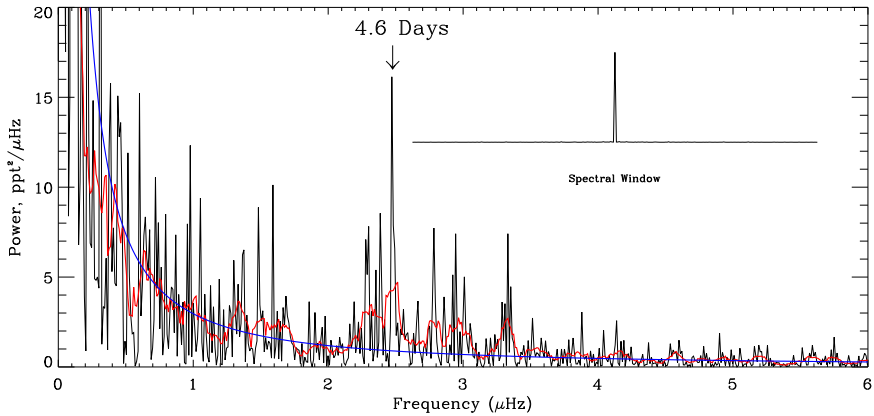}
      \caption{
		Power density spectrum with spectral window at the same frequency scale. Here the spectrum is shown in black, with a \hbox{9-bin} \hbox{($\approx$0.1 $\mu$Hz)}, moving-mean smoothed spectrum in red, and a fitted power-law background in blue.}
         \label{Fig:PowerSpectrum}
   \end{figure*}

Zooming in on a three-month section of the data between Dec 2003 and Mar 2004 (Fig. \ref{Fig:TimeDomain}, bottom) we can see clearly an oscillation with a period of $\approx$4.6 days and an amplitude of a few ppt. A power density spectrum of the time series is shown in Figure \ref{Fig:PowerSpectrum}. A prominent spike and concentration of power can be seen at a frequency of $\approx$2.5 $\mu$Hz, in agreement with the strong oscillation of 4.6 days visible in the time series. Further spikes and concentrations of power can be seen within the range 2 to 4 $\mu$Hz; in particular, the eye is drawn to three prominent features, located between approximately 2.2 to 2.6 $\mu$Hz (including the above noted spike), 2.7~to~3.1~$\mu$Hz, and 3.2~to~3.4~$\mu$Hz. In the smoothed spectrum (red), the features at 2.2~to~2.6~$\mu$Hz and 2.7~to~3.1~$\mu$Hz could be construed as being composed of two or more distinct concentrations of power. 

Missing data in a time series can result in power being redistributed between frequencies. This redistribution is represented in a plot of the spectral window, shown in Figure \ref{Fig:PowerSpectrum} as an inset. Apart from the central peak, there are evidently no prominent spikes in the spectral window, which is very clean. Outside of the range visible in the plot, peaks occur at the diurnal frequency of 11.57 $\mu$Hz with a size of roughly one thousandth that of the central peak. Sub-harmonics of a day do not appear to be visible. To ensure redistribution of power by the window function was not in any way responsible for the excess power observed between 2 to 4 $\mu$Hz in the real power density spectrum, we created artificial data with a power-law background model -- as described in the first paragraph of the results section below -- to which the window function of the original data was applied. No such simulation gave any concentration of power in the region 2 to 4 $\mu$Hz of comparable prominence to what is seen in the $\beta$ UMi power density spectrum.

\section{Results}

Next, we applied statistical tests to determine whether the prominent features in the power density spectrum between 2~and~4~$\mu$Hz could be accounted for as a product of smooth broad-band background noise. In these statistical tests the power was compared to the local background, and any significant deviations noted. In the case of ${\rm \beta}$\,UMi, a power-law model of the form $a+b\nu^{-c}$, in which $a$, $b$, and $c$ are parameters to be fitted, was chosen to represent the background. To account for the redistribution of power by the spectral window the background model was convolved with the spectral window, during fitting. In this way we sought to ensure that the model background gave a good representation of the actual background.

Statistical tests, described in \citet{2007MNRAS.382L..48T}, were performed upon the power density spectrum, detecting first the presence of any significant spikes over the background noise, that is, power concentrated into a single bin. This may indicate the central frequency of a sharp mode. A second test highlighted any significant concentrations of power across a narrow frequency range. This may in turn indicate a broad mode in which the power is spread over a number of adjacent bins.

Considering the spectrum as a whole and setting a threshold of 1\%, only one instance of a single bin with a significant excess of power was observed, located at a frequency of \hbox{$\approx$ 2.46 $\mu$Hz}. This bin contains 18.5 times the background power at this frequency, giving a `false alarm' probability of lower than 0.1 \%. Considering power within $\approx$0.1~$\mu$Hz ranges, three concentrations were highlighted as having less than a 1\% chance of being a product of noise. These ranges were centred at approximately 2.33, 2.44, and 2.98 $\mu$Hz.

These results suggest that the region of excess power between 2~and~4~$\mu$Hz has some structure to it. We first tested the null-hypothesis that the power between 2~and~4~$\mu$Hz is represented by a single resonant peak against the hypothesis that it would be reproduced better by two peaks. This was done by fitting resonant peak profiles (by the method described below) reflecting each hypothesis, and performing a likelihood ratio test \citep{1995ASPC...76..314A}. This revealed that fitting two resonant profiles represented a significant increase in the quality of fit over a single profile.

Next we considered each of the two broad concentrations of power, between 2.2 to 2.6 $\mu$Hz and 2.7~to~3.1~$\mu$Hz, respectively. We tested whether the structure of each concentration could be ascribed to the effects of noise upon a single broad mode or represent two separate narrow modes. In this case it was found that fitting two resonant profiles offered an insignificant increase in quality-of-fit over a single profile. We therefore treated each of the two concentrations of power as composed of a single broad mode.

\begin{table}

\centering

\caption{ Best-fitting estimates of identified modes. }

\label{Table:fits}

\begin{tabular}{cccc}

\hline \hline
Frequency	&	Width \textbf{(FWHM, $\Delta$)}			&	Height				&	RMS amplitude		\\
($\mu$Hz)	&	($\mu$Hz)	&	[(ppt)$^2/\mu$Hz]	&	(ppt)		\\
\hline
	$2.44 \pm 0.04$	&	$0.2 \pm 0.1$	&	$5.4 \pm 2.2$	&	$1.3 \pm 0.4 $	\\
	$2.92 \pm 0.05$	&	$0.2 \pm 0.1$	&	$2.8 \pm 1.1$	&	$0.9 \pm 0.3 $	\\
\hline \hline
\end{tabular}

\end{table}

\begin{figure}
   \centering
   \includegraphics[width=0.4\textwidth]{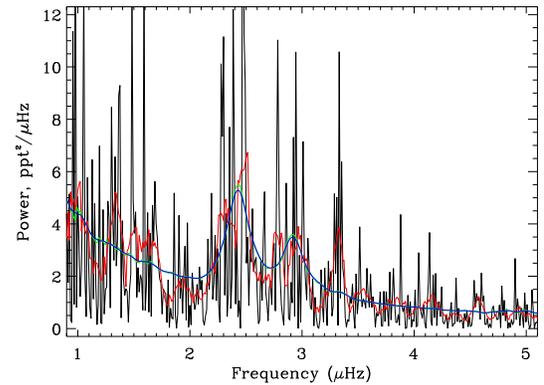}
      \caption{ Fits to the most prominent features in the power density spectrum. Here the raw spectrum is show in black, and a \hbox{9-bin} \hbox{($\approx$0.1 $\mu$Hz)} moving-mean smoothed spectrum in red. The raw fitting is shown in green, and smoothed over 9 bins in blue.}         
		\label{Fig:Fitting}
   \end{figure}

A fitting to the features in the power density spectrum is shown in Figure \ref{Fig:Fitting} in which each of the two concentrations of power at 2.2~to~2.6~$\mu$Hz and 2.7~to~3.1~$\mu$Hz was fitted as a single mode by means of a maximum likelihood technique. As the mode lifetimes appear not to be significantly longer than the period of the oscillations, it was necessary in this case to use the full resonant profile describing a classically damped oscillator as the model to describe each mode, instead of the usual simplification of a Lorentzian model. This fitting was performed by simultaneously fitting to the spectrum a power-law model of the background and the two resonant profiles, which were constrained to have the same width. The window function can alter the parameters describing a fitted mode, so we convolved our model with the window function. By this means we hoped to fit a true description of the modes better. The best-fitting parameters describing each mode are listed in Table \ref{Table:fits}.

As the signal-to-noise ratio is low, more simulations were performed to further ensure that the fitted parameters were a robust representation of the power density spectrum. In these a model spectrum was made consisting of the background and two modes described by the parameters in Table \ref{Table:fits}. Artificial power density spectra, $P^\prime (\nu_i)$, were then created point-wise in which $P^\prime (\nu_i) = -\ln{(x_i)} P(\nu_i)$, where $x_i$ is a uniformly distributed random variable in the range $[0,1]$, and $P(\nu_i)$ is the model power at the frequency $\nu_i$. By this method, the artificial spectra showed the correct $\chi^2$ with two degrees of freedom statistics expected for power spectra \citep{1990ApJ...364..699A}. Fittings to these artificial spectra were performed and the returned parameters compared to those of the input model. Each fitted parameter showed a distribution around a mean value consistent with the initial fitting. The standard deviation of each parameter was used to determine the error estimates shown in Table \ref{Table:fits}.

Considering other spikes and concentrations of power in the region of the two identified modes, it is tempting to speculate that some of these may represent additional modes. However, we note the strongest feature at $\approx$3.3~$\mu$Hz occurs at the same location as an alias introduced by the window function, as can be observed by the slight hump seen in the fitting in Figure \ref{Fig:Fitting}. In addition, the feature at $3.3~\mu$Hz does not show the width of the identified modes at 2.44 and 2.92~$\mu$Hz, and occurs at a lower frequency than would be expected for a higher harmonic respecting the same frequency spacing shown between the other two modes.

\section{Discussion}

Two features have been identified as unlikely to be part of the background in both probabilistic and numerical models of the data. We suggest that these features are radial modes separated by the large frequency spacing. Given the quality of the data, we feel it is not possible to say whether there might also be non-radial modes present. However, we believe that we can exclude the possibility of the separation between the features being several times the large frequency spacing, as this would require modes to be missing from the spectrum, and the lower spacing thus implied would give an unrealistic mass estimate (see below).

As we have measured two independent seismic quantities -- the frequency of maximum power, $\nu_{\rm max}$, and large frequency spacing, $\Delta\nu$ -- we may use Eqs. 9 and 10 in \citet{1995A&A...293...87K} to estimate the asteroseismic mass:
 \begin{equation}
 \label{eqn:mass}
 \left( \frac{M}{\rm M_{\odot}} \right) = \frac{(\nu_{\rm max}/3050\ \mu\mathrm  
 {Hz})^3(T_{\rm eff}/5777\ {\rm K})^{3/2}}{(\Delta\nu/134.9\ \mu{\rm Hz})^4},
 \end{equation}
where 3050~$\mu$Hz, 134.9~$\mu$Hz, and 5777~K are respectively the frequency of maximum power, large spacing, and effective temperature of the Sun (errors on these values will be insignificant when compared with errors on the $\beta$ UMi values and have been ignored). 

We determined the frequency of maximum power by smoothing the power spectrum by means of a boxcar (moving mean) smoothing filter of width $2\Delta\nu$ applied twice. This will produce a single hump of excess power that is insensitive to the discrete peaks of the oscillation spectrum.
The frequency of maximum power observed with this method was \hbox{$\nu_{\rm max} = 2.61\pm 0.08~\mu$Hz}, where the uncertainty comes from the scatter on results observed for the artificial spectra described in Section 3. We note that this location seems somewhat inconsistent with predictions based upon the scaling laws (Table \ref{Table:BetaUmiParams}).

When we use the observed values of \hbox{$\Delta\nu = 0.48\pm 0.06 ~\mu$Hz} and \hbox{$\nu_{\rm max} = 2.61\pm 0.08~\mu$Hz} we obtain a mass estimate  of $2.3\pm 1.4~{\rm M}_{\odot }$. This means of estimating the mass is highly sensitive to both $\nu_{\rm max}$ and $\Delta\nu$ (third and fourth power dependence, respectively). As both values have significant errors, this leads to a very large fractional uncertainty in the derived mass. We have therefore used two other approaches to estimate the mass, which give better constrained estimates. In the first approach \citep{2008ApJ...674L..53S}, we use estimates of the luminosity (from use of the parallax) and effective temperature from Table 1, in combination with the value for $\nu_{\rm max}$. This expression is -
 \begin{equation}
 \left( \frac{M}{\rm M_{\odot}} \right) = \frac{ (\nu_{\rm max}/3050\ \mu\mathrm  
 {Hz}) (L/{\rm L_{\odot}) } } { (T_{\rm eff}/5777\ {\rm K})^{3.5} }.
 \end{equation}
For $\beta$ UMi this method returns a mass estimate of $1.4 \pm 0.2\ {\rm M}_{\odot}$. In the second approach, we make use of the dependence of $\Delta\nu$ on the square root of the mean density of the star. This expression is:
 \begin{equation}
 \left( M / \rm M_{\odot} \right) =  (\Delta \nu/134.9\ \mu\mathrm{Hz})^2 (R/{\rm   
 R_{\odot})^3 }.
 \end{equation}
When we use the independent interferometric determination of \hbox{$R = 44.4\pm 0.7~{\rm R}_{\odot}$} (Table 1), we obtain a mass of \hbox{$1.1\pm 0.3~{\rm M}_{\odot}$}. The mass estimates obtained from these two approaches are consistent with each other, but do differ significantly from the $\log g$ and evolutionary track mass estimates shown in Table \ref{Table:BetaUmiParams}. Since the two approaches make use of input data that are independent, we may give a combined mass estimate of $1.3\pm 0.3~{\rm M}_{\odot}$.

From the fitted width, a mode lifetime $\tau = 1/(\pi\Delta)$ (where $\Delta$ is the mode width) of $18 \pm 9$~days is obtained giving a low Q-factor ($\nu/\Delta$) of approximately 12. It has been suggested that Sun-like oscillations may show a trend of decreasing quality with decreasing frequency. Our results here respect this trend, when considered in the context of other results obtained for $\xi$ Hya \citep[$\nu_{\rm max} \approx 80~\mu$Hz, $Q \approx 55$; ][]{2006A&A...448..709S} and Arcturus  \citep[$\nu_{\rm max} = 3.5~\mu{\rm Hz}, Q \approx 20$; ][]{2007MNRAS.382L..48T}. 

We observe a maximum RMS mode amplitude of \hbox{$1.3\,\pm\,0.4$~ppt}. This observed amplitude agrees with our original prediction of \hbox{$1.6\,\pm\,0.5$~ppt} shown in Table \ref{Table:BetaUmiParams}.
However, the Table \ref{Table:BetaUmiParams} prediction was made from the scaling laws based upon an estimated mass of \hbox{$2.2\ {\rm M}_{\odot }$}. The prediction assumes that mode amplitudes scale linearly with the luminosity-to-mass ratio ($L/M$). If we re-calculate the predicted amplitude using our estimated mass from above \hbox{($1.3 \pm 0.3 \ {\rm M}_{\odot }$)}, we obtain a value of \hbox{$2.8\pm 0.8$~ppt}. While this prediction is more than double our observed amplitude, the large errors on both the prediction and observation mean the two values lie only 1.5$\sigma$ apart.

%______________________________________________________________________

\section{Conclusions}

Variability of period $\approx$4.6 days has been observed in the K4III giant star, ${\rm \beta}$\,UMi. Two modes appear to be present at 2.44 and 2.92\,$\mu$Hz, with a spacing between harmonics of 0.48~$\mu$Hz. The mode lifetime is estimated to be \hbox{$18\pm 9$ days.}

Derived values for the frequency of maximum mode power and spacing between harmonics suggest that previous estimates of the mass, which have been in the range of $\approx 2.0\ {\rm to}\ 2.5\ {\rm M}_{\odot }$, may represent an overestimate of the true mass, so we tentatively suggest a lower mass for the star of $1.3 \pm 0.3\ {\rm M}_{\odot}$.

The maximum RMS mode amplitude is consistent with the predictions from scaling relations, falling about 1.5 $\sigma$ below predictions based on the revised mass estimate given above.

%______________________________________________________________________

\begin{acknowledgements}

The authors acknowledge the support of STFC. SAS also acknowledges the support of the School of Physics and Astronomy, University of Birmingham. SMEI was designed and constructed by USAFRL, UCSD, Boston College, Boston University, and the University of Birmingham. The authors also thank the anonymous referee for her/his useful comments. 
      
\end{acknowledgements}

%______________________________________________________________________

% bibliography 
\bibliographystyle{aa} % style aa.bst
\bibliography{9738} % your references Yourfile.bib

\newpage

\end{document}